\documentclass[]{iopart}
\usepackage[english]{babel}
\usepackage[utf8x]{inputenc}
\usepackage{iopams}
\expandafter\let\csname equation*\endcsname\relax
\expandafter\let\csname endequation*\endcsname\relax
\usepackage{amsmath}
\usepackage{graphicx}
\usepackage{hyperref}
\usepackage{cite}
\usepackage{xifthen}
\usepackage{xfrac}

\begin{document}
  \title{Bosonic Hofstadter butterflies in synthetic antiferromagnetic patterns}

  \author{Yury S Krivosenko$^1$, Ivan V Iorsh$^1$ and Ivan A Shelykh$^{1, 2}$}
  \address{$^1$ Department of Physics and Engineering, ITMO University, St.\,Petersburg 197101, Russia}
  \ead{y.krivosenko@gmail.com}
  \address{$^2$ Science Institute, University of Iceland, Dunhagi 3, IS-107, Reykjavik, Iceland}

  \begin{abstract}
    The emergence of Hofstadter butterflies for bosons in synthetic-gauge-field antiferromagnetic patterns is theoretically studied.
    We report on a specific tight-binding model of artificial antiferromagnetic structures incorporating both nearest and next-to-nearest neighbour tunnelings and allowing for the formation of the fractal spectra even with the vanishing gauge field flux through the lattice.
    The model is applied to square and honeycomb lattices.
    Possible experimental realization is suggested for the lattices of microring resonators connected by waveguides.
    Finally, the structure of the butterflies is analyzed for different points in the magnetic Brillouin zone for both the ferromagnetic and antiferromagnetic patterns.
  \end{abstract}

  \vspace{2pc}



  \renewcommand{\vec}[1]{\mathbf{#1}}
\newcommand{\sigmai}[1]{\widehat{\sigma}_{#1}}
\newcommand{\sigmax}{\sigmai{x}}
\newcommand{\sigmay}{\sigmai{y}}
\newcommand{\sigmaz}{\sigmai{z}}
\newcommand{\sigmao}{\sigmai{0}}
\newcommand{\atob}{
\begin{pmatrix}
  0 & 0 \\
  1 & 0
\end{pmatrix}
}
\newcommand{\btoa}{
\begin{pmatrix}
  0 & 1 \\
  0 & 0
\end{pmatrix}
}
\renewcommand{\e}{\mathrm{e}}
\newcommand{\abs}[1]{\left| #1 \right|}
\newcommand{\pars}[1]{\left( {#1} \right)} 
\newcommand{\sqbr}[1]{\left[ {#1} \right]} 
\newcommand{\hc}{\text{h.c.}}
\newcommand{\transp}{\mathtt{T}}

\newcommand{\sumnn}{\sum\limits_{n_1, n_2}}
\newcommand{\sumkk}{\sum\limits_{k_1, k_2}}
\newcommand{\instate}{\langle n_1, n_2 |}
\newcommand{\tnn}{t_{\text{1}}}
\newcommand{\tnnn}{t_{\text{2}}}
\newcommand{\ham}{\widehat{\mathcal{H}}}
\newcommand{\Hhc}[1][]{\ifthenelse{\isempty{#1}}{\ham_{6}}{\ham_{6{#1}}}}
\newcommand{\Hsq}[1][]{\ifthenelse{\isempty{#1}}{\ham_{4}}{\ham_{4{#1}}}}
\newcommand{\Hhcold}[1][]{\ifthenelse{\isempty{#1}}{\widehat{H}_{6}}{\widehat{H}_{6{#1}}}}
\newcommand{\Hsqold}[1][]{\ifthenelse{\isempty{#1}}{\widehat{H}_{4}}{\widehat{H}_{4{#1}}}}
\newcommand{\Hhcnn}{\Hhcold[\text{nn}]}
\newcommand{\Hhcnnn}{\Hhc[\text{nnn}]}
\newcommand{\Hsqnn}{\Hsqold[\text{nn}]}
\newcommand{\Hsqnnn}{\Hsq[\text{nnn}]}
\newcommand{\Hhck}{\Hhc[\vec{k}]}
\newcommand{\Hsqk}{\Hsq[\vec{k}]}
\newcommand{\Hhckone}{\Hhc[\text{k1}]}
\newcommand{\Hsqkone}{\Hsq[\text{k1}]}
\newcommand{\Hzero}{\widehat{H}_{\text{aux}}}
\newcommand{\Hbulk}{\ham_{\text{bulk}}}
\newcommand{\Hqq}{\widehat{\mathcal{U}}_{m}}
\newcommand{\Jqq}{\widehat{\mathcal{J}}_{m}}
\newcommand{\rot}{\mathop{\mathrm{curl}}\nolimits}
\newcommand{\diag}[1]{\mathop{\mathrm{diag}}\nolimits \left( {#1} \right)}
\newcommand{\slA}{\mathcal{A}}
\newcommand{\slB}{\mathcal{B}}
\newcommand{\kstate}{| k_1, k_2 \rangle}
\newcommand{\kstateconj}{\langle k_1, k_2 |}
\newcommand{\kappastate}{| \varkappa_1, \varkappa_2 \rangle}
\newcommand{\kappastateconj}{\langle \varkappa_1, \varkappa_2 |}
\newcommand{\const}{\mathrm{const}}
\newcommand{\ipix}{i\pi\xi}
\newcommand{\kprojector}{| k_2 \rangle \langle k_2 |}
\newcommand{\hcl}{\text{hc}}
\newcommand{\sql}{\text{sq}}
\newcommand{\nnew}{n_1'}
\newcommand{\iq}{\widehat{I}_Q}
\newcommand{\ip}[1]{\widehat{I}'_{#1}}
\newcommand{\ipp}[1]{\widehat{I}''_{#1}}
\newcommand{\ippt}[1]{\widehat{I}''^{\mathsf{T}}_{#1}}
\newcommand{\cq}[1][]{\ifthenelse{\isempty{#1}}{\widehat{\mathcal{C}}}{\widehat{\mathcal{C}}^{#1}}}
\newcommand{\dq}[1][]{\ifthenelse{\isempty{#1}}{\widehat{\mathcal{D}}}{\widehat{\mathcal{D}}^{#1}}}
\newcommand{\fq}[1][]{\ifthenelse{\isempty{#1}}{\widehat{\mathcal{F}}}{\widehat{\mathcal{F}}^{#1}}}
\newcommand{\gq}[1][]{\ifthenelse{\isempty{#1}}{\widehat{\mathcal{G}}}{\widehat{\mathcal{G}}^{#1}}}
\newcommand{\OO}{\widehat{\mathcal{O}}}
\newcommand{\thetaraman}{\theta_{\text{R}}}
\newcommand{\diff}{\mathrm{d}}
\newcommand{\bigo}[1]{\mathop{\mathcal{O}}\nolimits \left( {#1} \right)}
\newcommand{\mpoint}{\widetilde{M}}

  \begin{section}{Introduction}
  The fractal electron spectrum originating in a two-dimensional gas of electrons on a lattice subject to homogeneous magnetic field was first described in~\cite{Hofstadter1976} by Hofstadter. Due to the characteristic shape of the spectrum, the effect was later called
  Hofstadter butterfly (HB). Since then HBs have been revealed in a variety of systems, ranging from electrons in 2D lattices \cite{Gumbs1997, Oh2000, Oh2001, Li2011} to the systems of trapped cold atoms  \cite{Yilmaz2015} and exciton polaritons \cite{Banerjee2018}. The latter two systems consist of electrically neutral particles, and thus to show the fractal spectra instead of the real magnetic field require synthetic gauge fields \cite{Celi2014,Banerjee2018}.

  There is a plethora of the physical effects related to the fractal nature of the spectrum. Padavi\'{c} \textsl{et al.} \cite{Padavic2018} reported on occurence of HB in topological phase diagram of Su-Schrieffer-Heeger ladder.
  Du \textsl{et al.} \cite{Du2018} applied Floquet theory and thus examined the influence of monochromatic field on HB in kagome and triangular lattices. Duncan \textsl{et al.} \cite{Duncan2020} researched topological modes in quasicrystals and observed HBs as well. The authors of \cite{Colella2019} gave an account of HB in square lattices with a synthetic magnetic field modified by external pump. Hafezi \textsl{et al.} designed the artificial gauge field in the square lattice of microring resonators connected by the waveguides. Also, HB-like spectra were achieved in the system of microring resonators arranged in circle \cite{Zimmerling2020}. Otaki and Fukui \cite{Otaki2019} examined generalized two-dimensional Su-Schrieffer-Heeger model as an example of high-order topological insulators and described the appearance of HB-type spectra. Jaksch and Zoller \cite{Jaksch2003} suggested the Raman-laser-assisted tunneling as the tool for creating gauge fields and hence the butterfly for neutral atoms in optical lattices, and Aidelsburger and co-authors \cite{Aidelsburger2013} followed the scheme.

  The general hallmark of the above-mentioned investigations consists in the presence of non-zero flux of either real magnetic or artificial gauge filed (hereafter, we refer to the flux as to the \textit{magnetic flux}) through the lattice unit cell. Usually, for the HB to form, the ratio of magnetic flux over the magnetic flux quantum ($hc/e$) should equal a rational number, i.e. the ratio of an integer and a natural number, $p/q$. In the present paper, we perform an attempt to reach the HB-type spectra in synthetic antiferromagnetic (AFM) structures realized on square and honeycomb lattices with both nearest-neighbour (NN) and next-to-nearest-neighbour (NNN) interactions. In this case, HBs can still be observed even for the case of the vanishing net flux through elementary cell.  This result suggests, that the HB phenomena can be observed in a much wider class of the systems as it was believed previously, and pave the way to the search of novel materials characterized by the fractal spectra.

  The paper is organized as follows. In Sec.~\ref{sec:theory}, the generic theoretical model is introduced and a scheme of experimental implementation is proposed. Then in Sec.~\ref{sec:rnd}, the main results are presented and analysed. The main findings are summarized in Sec.~\ref{sec:conclusion}.
\end{section}

  \begin{section}{Theory \label{sec:theory}}
  \begin{subsection}{Basic concepts, real space}
    \begin{figure}[t]
      \centering
      \includegraphics[height=3.5cm]{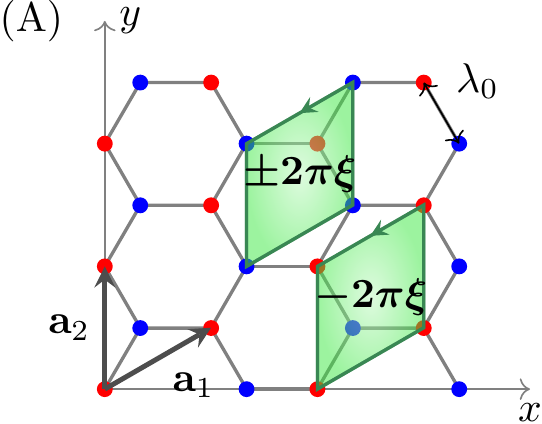}
      \hspace{1cm}
      \includegraphics[height=3.5cm]{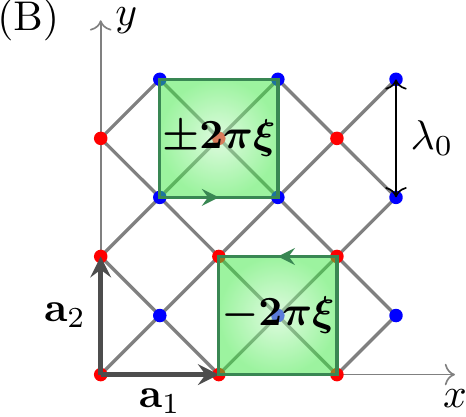}
      \caption{The (A) honeycomb and (B) square lattices. $\vec{a}_1$ and $\vec{a}_2$ are the translation vectors, red and blue dots are the sites belonging to sublattices $\slA$ and $\slB$, respectively. Only the nearest neighbour bonds are displayed in light gray. The green areas designate the unit plaquettes in each sublattice, $-2\pi\xi$ and $\pm 2\pi\xi$ values show the phase gained after committing a circular path around each plaquette. The choice of $\pm$ sign in the upper green areas in both (A) and (B) panels gives the AFM ($+2\pi\xi$) and FM ($-2\pi\xi$) patterns.}
      \label{fig:lattices}
    \end{figure}

    We start with the bare Hamiltonians of the honeycomb ($\Hhc$) and square ($\Hsq$) lattices with two sites per unit cell. Both nearest neighbour and next-to-nearest neighbour hoppings are included, the corresponding amlitudes are $\tnn$ and $\tnnn$, respectively. Neither magnetic nor other gauge field is yet present, polarization degree of freedom is not taken into account.
    Firstly, consider the following auxiliary operator:
    \begin{align}
      \Hzero &= \tnn \sumnn \left[
      | n_1, n_2 \rangle \instate \otimes \atob \right. \nonumber \\ &+
      | n_1+1, n_2 \rangle \instate \otimes \btoa +
      \left.
      | n_1, n_2+1 \rangle \instate \otimes \btoa \right] \nonumber \\ &+
      \tnnn \sumnn \biggl[
      | n_1+1, n_2 \rangle \instate \otimes \sigmao+
      | n_1, n_2+1 \rangle \instate \otimes \sigmao \biggr]
      + \hc,
      \label{eq:Hzero}
    \end{align}
    {The operator is written in the basis of Wannier states $| n_1, n_2 \rangle \otimes | \alpha \rangle$ localized at the lattice sites. $n_1$ and $n_2$ signify the coordinates of a unit cell along the crystallographic directions $\vec{a}_1$ and $\vec{a}_2$, see Fig.~\ref{fig:lattices} ($n_1 = 0, 1, \ldots (N_1{-}1)$, $n_2 = 0, 1, \ldots (N_2{-}1)$).}
    $|\alpha\rangle$ designates the inner-cell state ($\alpha = A, B$).
    Tensor product serves to separate the external (cell-position, $| n_1, n_2 \rangle$) and internal (inner-cell, $| \alpha \rangle$) states.
    The $2{\times}2$ matrices and $\sigmao$ (the identity matrix) act in the inner-cell states subspace and are thus two-dimensional.
    For the periodic boundary conditions, the terms $n_{1(2)} {+} 1$ are taken modulo $N_{1(2)}$.

    Then,
    \begin{subequations}{\label{eq:bare_hamiltonians}}
      \begin{align}
        \Hhcold &= \Hzero + \tnnn \sumnn \biggl[ | n_1, n_2+1 \rangle \langle n_1+1, n_2 | \otimes \sigmao + \hc \biggr]
        \label{eq:Hhc}
        \intertext{and}
        \Hsqold &= \Hzero + \tnn \sumnn \biggl[ | n_1+1, n_2+1 \rangle \instate \otimes \btoa + \hc \biggr]
        \label{eq:Hsq}
      \end{align}
    \end{subequations}
    Despite the fact that square lattice Hamiltonian can be constructed even with a single site per unit cell, we introduce here the two-sites unit cell for the sake of uniformity.

    In this paper, we propose to design the FM and AFM patterns in compliance with the following scheme. First, within the tight-binding approximation, the influence of the magnetic field on the behaviour of spinless charged particles is manifested in the occurrence of the phase factor $\exp{\left( -i \theta(s_1, s_2) \right)}$ at the hopping amplitudes, the so-called Peierls substitution. For hopping from site $s_1$ to $s_2$, the phase $\theta(s_1, s_2)$ equals the linear integral \cite{Yilmaz2015}
    \begin{equation}{\label{eq:theta_s1s2}}
      \theta(s_1, s_2) = \frac{2\pi}{\phi_0} \int\limits_{s_1}^{s_2} \vec{A} {\cdot} \mathrm{d} \vec{l} =
      2\pi \cdot \frac{\Phi}{\phi_0} \cdot
      \dfrac{\int\limits_{s_1}^{s_2} \vec{A} {\cdot} \mathrm{d} \vec{l}}{\Phi},
    \end{equation}
    where $\nabla\times\vec{A} = \vec{B}$ is the effective magnetic field corresponding to the artificial gauge field, $\Phi$ is its flux through the unit cell, and $\phi_0$ is the quantum of magnetic field flux. We then utilize the fact that the honeycomb (square) lattice can be presented as two triangular (square) sublattices, $\slA$ and $\slB$, shifted with respect to each other (the red and blue sublattices in Fig.~\ref{fig:lattices}). Naturally,  NNN hoppings leave  the particle within the same sublattice, whereas NN hoppings correspond to the inter-sublattice process.
    Our main concept in constructing and distinguishing the FM and AFM phases consists in the assumption that the sublattices can be subject to the different gauge fields: the magnetic field applied to sublattice $\slA$ is homogeneous, directed along $z$-axis and equals $B_0$, whereas it is $\pm B_0$ for the sublattice $\slB$ in the FM ($+$) and AFM ($-$) phases. Moreover, the NN hopping matrix elements are assumed to be purely real and positive. The proposal for realization and justification of such a gauge field is presented in Subsection~\ref{sec:exp_realization}.

    Using Landau gauge, i.e. $\vec{A} = (0, xB, 0)$, one can derive the linear integral in \eqref{eq:theta_s1s2} and hence the phases $\theta(s_1, s_2)$ \cite{Gumbs1997, Oh2001, Oh2000}. The former is
    \begin{equation}{\label{eq:intAdl_Landaugauge}}
      \int\limits_{s_1}^{s_2} \vec{A} {\cdot} \diff \vec{l} = B \, \sin{\varphi} \sqbr{x(n_1, n_2, \tau) \lambda' + \cos{\varphi} \frac{\lambda'^2}{2}}.
    \end{equation}
    The integral is evaluated for hopping from the site specifed by the coordinates $n_1$, $n_2$, and $\tau$ to its next-to-nearest neighbour. $\tau = 0(1)$ for $\slA(\slB)$ sublattice, $x(n_1, n_2, \tau)$ is the $x$-coordinate of the initial site, $\varphi$ is the angle between the hopping direction and $x$-axis, $\lambda'$ is hopping distance. For the honeycomb lattice $\lambda' = \lambda_0 \sqrt{3}$, and for the square one $\lambda_0$, see Fig.~\ref{fig:lattices}. The magnetic field $B$ equals $B_0$ for the both sublattices in the FM phase, and $\pm B_0$ for sublattice $\slA(+)$ and $\slB(-)$ in the AFM phase.

    We arrange the coordinate axes so as the $y$-axis direction coincides with that of translation unit vector $\vec{a}_2$.
    Hence, we obtain the initial positions as
    \begin{subequations}
      \begin{align}
        x_{6} (n_1, n_2, \tau) &= \frac{\lambda_0 \pars{3n_1 + \tau}}{2}, \\
        x_{4} (n_1, n_2, \tau) &= \lambda_0 \pars{n_1 + \frac{\tau}{2}}
      \end{align}
    \end{subequations}
    for the honeycomb and square lattices, respectively.

    As soon as the NN hopping amplitudes are supposed not to vary, we split each of the Hamiltonians \eqref{eq:bare_hamiltonians} into two parts responsible for NN and NNN hoppings and come to the Hamiltonians modified by the presence of the gauge field, $\ham_{4(6)}$:
    \begin{subequations}{\label{eq:HMRs}}
      \begin{align}
        \Hhc^{\pm} &= \Hhcnn + \tnnn \sumnn \left[
        | n_1 + 1, n_2 \rangle \instate \otimes \cq[\pm]_{n_1}
        + | n_1, n_2 + 1 \rangle \instate \otimes \dq[\pm]_{n_1}
        \right. \nonumber\\
        &+ \left. | n_1, n_2 + 1 \rangle \langle n_1+1, n_2 | \otimes \cq[\pm]_{n_1}
        + \hc \right]
        \label{eq:H6M}
        \intertext{and}
        \Hsq^{\pm} &=  \Hsqnn + \tnnn \sumnn \left[
        | n_1+1, n_2 \rangle \instate \otimes \sigmao
        + | n_1, n_2+1 \rangle \instate \otimes \fq[\pm]_{n_1}
        + \hc \right],
        \label{eq:H4M}
      \end{align}
    \end{subequations}
    where $\cq$, $\dq$, and $\fq$ are the $2{\times}2$ diagonal operators dependent on the magnetic field and pattern:
    \begin{subequations}{\label{eq:2by2_magnetic}}
      \begin{align}
        \cq[\pm]_{n} &=
        \begin{pmatrix}
          \e^{-\ipix (n+\sfrac{1}{2})} & 0 \\
          0 & \e^{\mp \ipix (n + \sfrac{5}{6})}
        \end{pmatrix}, \\
        \dq[\pm]_{n} &=
        \begin{pmatrix}
          \e^{-2 \ipix n} & 0 \\
          0 & \e^{\mp 2 \ipix (n + \sfrac{1}{3})}
        \end{pmatrix}, \\
        \fq[\pm]_{n} &=
        \begin{pmatrix}
          \e^{-2\ipix n} & 0 \\
          0 & \e^{\mp 2 \ipix (n + \sfrac{1}{2})}
        \end{pmatrix}.
      \end{align}
    \end{subequations}
    The $+(-)$ superscript index denotes the FM (AFM) pattern revealed in the sign of the phase of the second diagonal terms. $\xi$ is the ratio $\Phi/\phi_0$. Below, the $\pm$ superscripts are placed at the Hamiltonians and the $2{\times}2$ operators, if specifying the magnetic phase is crucial, and are omitted otherwise.
  \end{subsection}

  \begin{subsection}{Reciprocal space and Harper Hamiltonians}
    As soon as the gauge phases in eqs.~\eqref{eq:HMRs} and \eqref{eq:2by2_magnetic} do not depend on $n_2$, the Fourier transform of the Hamiltonians along this direction can be straightforwardly performed.
    At this stage, the localized state $| n_1, n_2 \rangle$ is presented as
    \begin{equation}
      | n_1, n_2 \rangle = \frac{1}{2\pi} \sum\limits_{n_1, k_2} |k_2 \rangle {\otimes} |n_1\rangle \ \e^{-i n_2 k_2},
      \label{eq:n1n2_as_n1k2}
    \end{equation}
    where $k_2$ designates the corresponding wave number: it takes on the values of $2\pi m_2/N_2$ with $m_2$ listing the integers within the $[-N_2/2, N_2/2)$ range. Thus, collecting eqs. \eqref{eq:Hzero}, \eqref{eq:bare_hamiltonians}, \eqref{eq:HMRs}, and \eqref{eq:n1n2_as_n1k2}, we arrive at
    \begin{subequations}{\label{eq:Hmag_k1}}
      \begin{align}
        \Hhc &= \sum\limits_{k_2} \kprojector \otimes \Hhckone
        \nonumber
        \intertext{with}
        \Hhckone &= \sum\limits_{n_1}
        \left\{
        \tnn \left[
        | n_1 \rangle \langle n_1 | \otimes
        \begin{pmatrix}
          0 & \e^{-i k_2} \\
          1 & 0
        \end{pmatrix}
        + | n_1+1 \rangle \langle n_1 | \otimes
        \begin{pmatrix}
          0 & 1 \\
          0 & 0
        \end{pmatrix}
        \right]
        \right. \nonumber \\
        &+
        \tnnn \left[
        | n_1+1 \rangle \langle n_1| \otimes \cq_{n_1}
        + \e^{-i k_2} \, | n_1 \rangle \langle n_1| \otimes \dq_{n_1} \right.
        \nonumber \\
        &+ \left.\left. \e^{-i k_2} \, | n_1 \rangle \langle n_1+1| \otimes \cq_{n_1}
        \right] \right\}
        + \hc,
        \label{eq:H6M_k2}
      \end{align}
      and
      \begin{align}
        \Hsq  &= \sum\limits_{k_2} \kprojector \otimes \Hsqkone \nonumber
        \intertext{with}
        \Hsqkone &=
        \sum\limits_{n_1}
        \left\{ \tnn \left[
        | n_1 \rangle \langle n_1 | \otimes
        \begin{pmatrix}
          0 & \e^{-i k_2} \\
          1 & 0
        \end{pmatrix}
        + | n_1+1 \rangle \langle n_1 | \otimes
        \begin{pmatrix}
          0 & 1 + \e^{-i k_2} \\
          0 & 0
        \end{pmatrix}
        \right] \right. \nonumber \\
        &+ \tnnn \left. \left[
        | n_1 + 1 \rangle \langle n_1 | \otimes \sigmao +
        \e^{-i k_2} | n_1 \rangle \langle n_1 | \otimes \fq_{n_1}
        \right] \right\}
        + \hc
        \label{eq:H4M_k2}
      \end{align}
    \end{subequations}

    Henceforth, we constrain ourselves to rational magnetic fluxes: $\xi = \Phi/\phi_0 = p/q \in \mathbb{Q} \cap (0, 1)$, where $p$ and $q$ are coprime integers.
    Within the assumption, all the $2 {\times} 2$ operators in sums \eqref{eq:Hmag_k1} (the right-side terminal operators in each summand) become translationally invariant:
    \begin{equation}{\label{eq:CDF_tr_inv}}
      \gq_{n+Q} = \gq_{n},
    \end{equation}
    whereafter $\gq$ stands for $\cq$, $\dq$, or $\fq$, and the translation period $Q$ is $2q$ and $q$ for the honeycomb and square lattices, respectively.
    The Hamiltonians \eqref{eq:H6M_k2} and \eqref{eq:H4M_k2} thus acquire translational symmetry with respect to the shift by $N_1 Q$ along $\vec{a}_1$.
    Then, we extend the system along $\vec{a}_1$ so as it got $N_1 Q$ unit cells in this dimension. $n_1$ varies in the range $0, 1 \ldots (QN_1{-}1)$ and can be rewritten as $n_1 = \nnew Q + s_1$ where $\nnew = 0, 1, \ldots (N_1{-}1)$ and $s_1 = 0, 1, \ldots (Q{-}1)$, which eventually lead to $\gq_{n_1} = \gq_{s_1}$.
    The system begins to effectively possess of $N_1$ supercells each of which contains $Q$ original unit cells.

    The intermediate Hamiltonians expressed in terms of $\nnew$ and $k_2$, as well as the final explicit forms of the bulk Hamiltonians defined as
    \begin{equation}
      \ham_{6(4)} = \sum\limits_{k'_1, k_2}
      |k'_1, k_2\rangle \langle k'_1, k_2| \otimes
      \ham_{6(4)\text{bulk}}
    \end{equation}
    are omitted here due to their cumbersomeness and are presented in Appendix~\ref{sec:App_expilicHams}.
    What is to be mentioned about them, is that, initially, both bulk Hamiltonians can be represented as the sum
    \begin{equation}
      \ham_{\text{bulk}} = \sum\limits_{m} \alpha_{m} \, \Hqq \otimes \Jqq,
    \end{equation}
    where each $\alpha_{m}$ is a constant proportional to either $\tnn$ or $\tnnn$, $\Hqq$ and $\Jqq$ are $Q{\times}Q$ and $2{\times}2$ matrices, respectively. Operators $\Hqq$ and $\Jqq$ generally inherit the information on the intra-supercell hoppings and the internal degree of freedom of the initial unit cell, respectively. The inter-supercell tunneling is conventionally incorporated in $\e^{\pm i k'_1}$ and $\e^{\pm i k_2}$ factors.

    The Hamiltonians at the first magnetic Brillouin zone $\Gamma$-points $(k'_1 {=} k_2 {=} 0)$ pairwise coincide for FM and AFM patterns:
    \begin{subequations}{\label{eq:Gamma_coincidence}}
      \begin{gather}
        \Hhc[\text{bulk}]^{+} (\Gamma) = \Hhc[\text{bulk}]^{-} (\Gamma), \\
        \Hsq[\text{bulk}]^{+} (\Gamma) = \Hsq[\text{bulk}]^{-} (\Gamma),
      \end{gather}
    \end{subequations}
    which follows from the equality $\gq[+]_{n} {+} \left(\gq[+]_n\right)^{\dagger} = \gq[-]_{n} {+} \left(\gq[-]_n\right)^{\dagger}$ and the diagonal (complementary) manner of their inclusion into the Hamiltonians for the square (honeycomb) structures.

    Cosidering the FM and AFM square lattice bulk Hamiltonians as functions of quasi momentum $\vec{k} = (k'_1, k_2)$, one obtains that they produce the same Hofstadter butterflies in the following cases:
    \begin{subequations}{\label{eq:Hams_k_minusk}}
      \begin{gather}{\label{eq:HamsSQ_k_minusk}}
        \Hsq[\text{bulk}]^{+} (k'_1, k_2) \sim \Hsq[\text{bulk}]^{+} (\pm k'_1, \pm k_2), \\
        \Hsq[\text{bulk}]^{-} (k'_1, k_2) \sim \Hsq[\text{bulk}]^{-} (\pm k'_1, \pm k_2)
      \end{gather}
      for all the combinations of $+$ and $-$ signs.
      Contrarily, the honeycomb lattice Hamiltonians generally differ.
    \end{subequations}
  \end{subsection}

  \begin{subsection}{Experimental realization}{\label{sec:exp_realization}}
    The realization we propose, see Fig.~\ref{fig:realization}, is based on the paper by Hafezi \textsl{et al.} \cite{Hafezi2011}. Each site here is a microring resonator. The resonators (red and blue rounds) are connected by the waveguides (red, blue, and gray solid elliptic lines). The phase acquirement can be achieved here, e.g., by tuning their relative lengths, see \cite{Hafezi2011}. The microring resonatros generally host photons of two polarizations, clockwise and counter-clockwise, thus introducing the pseudospin into the system. In the absence of specific scatterers, the pseudospin components can be considered as uncoupled thus separating the system into two independent subsystems. Here, we choose the counter-clockwise polarization (explicitly shown in the figure) and demonstrate only the square AFM pattern.

    To arrange the waveguides, we utilize three nominal layers: two layers to internally connect the sites within sublattice $\slA$ and sublattice $\slB$ (the red and blue waveguides), and another one to link the nearest neighbours (the gray ones). The lengths of $\slA$($\slB$)-waveguides are designed to result in $-(+)2\pi\xi$ flux through the square plaquette presented by the green area. The effective phase shift equals $0$ for the gray waveguides (nearest neighbour hoppings).

    \begin{figure}[t]
      \centering
      \includegraphics[scale=.6]{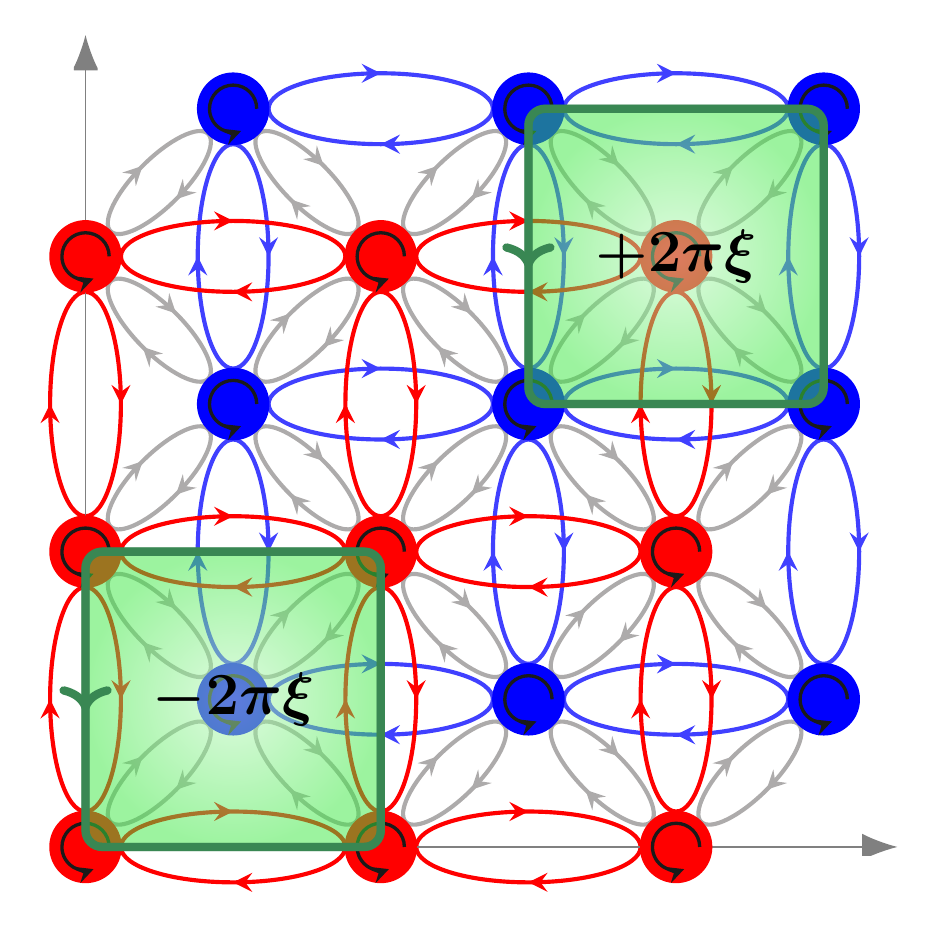}
      \caption{The square AFM structure. The red (blue) rounds and elliptically shaped curves display $\slA$ ($\slB$) sublattice sites and NNN connection waveguides, respectively. The gray curves depict the waveguides linking nearest neighbours. The green areas indicate unit plaquettes in both $\slA$ and $\slB$ sublattices with the corresponding phases, $\pm 2 \pi \xi$, equal negative the gauge field flux through the relative plaquette. The difference in waveguides lengths is not expicitly shown.}
      \label{fig:realization}
    \end{figure}

  \end{subsection}
\end{section}

  \begin{section}{Results and discussion \label{sec:rnd}}
  \begin{figure*}[t]
    \centering
    \includegraphics[width=1.0\textwidth]{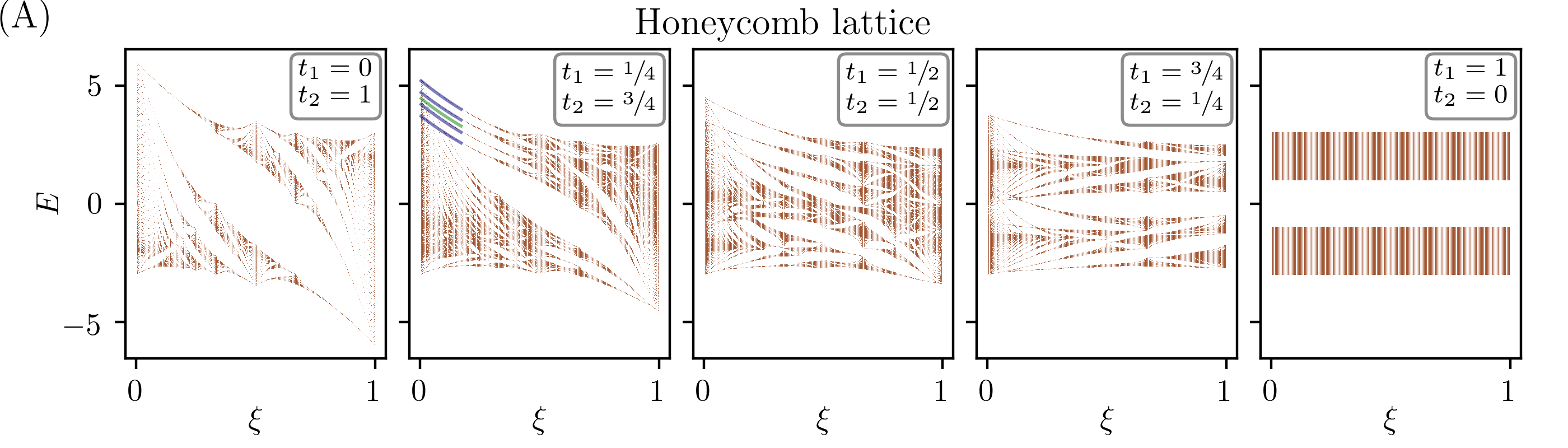}
    \includegraphics[width=1.0\textwidth]{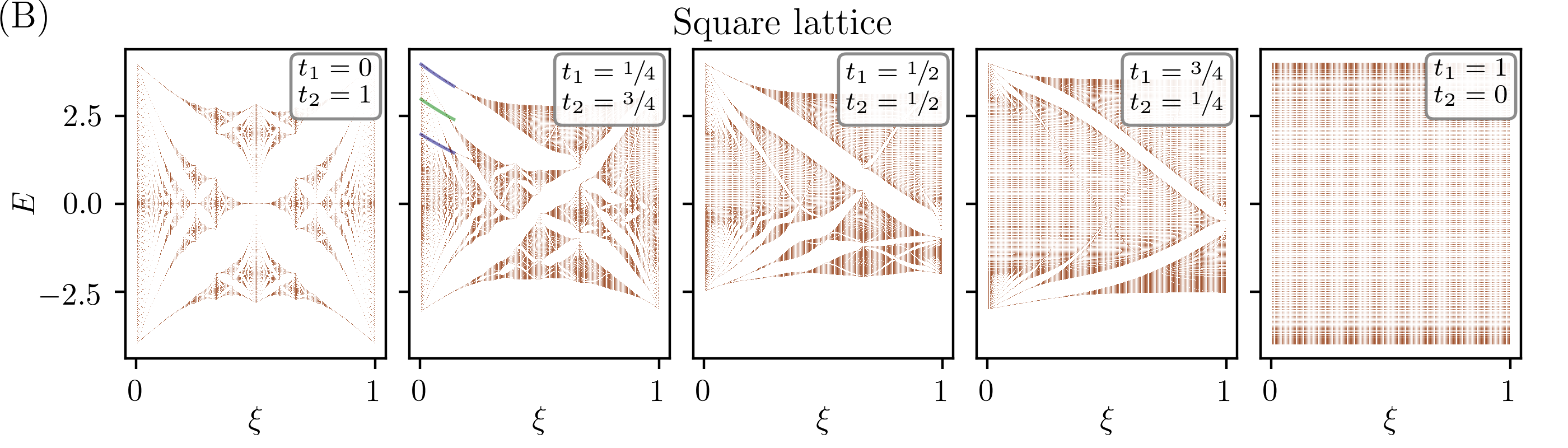}
    \includegraphics[width=1.0\textwidth]{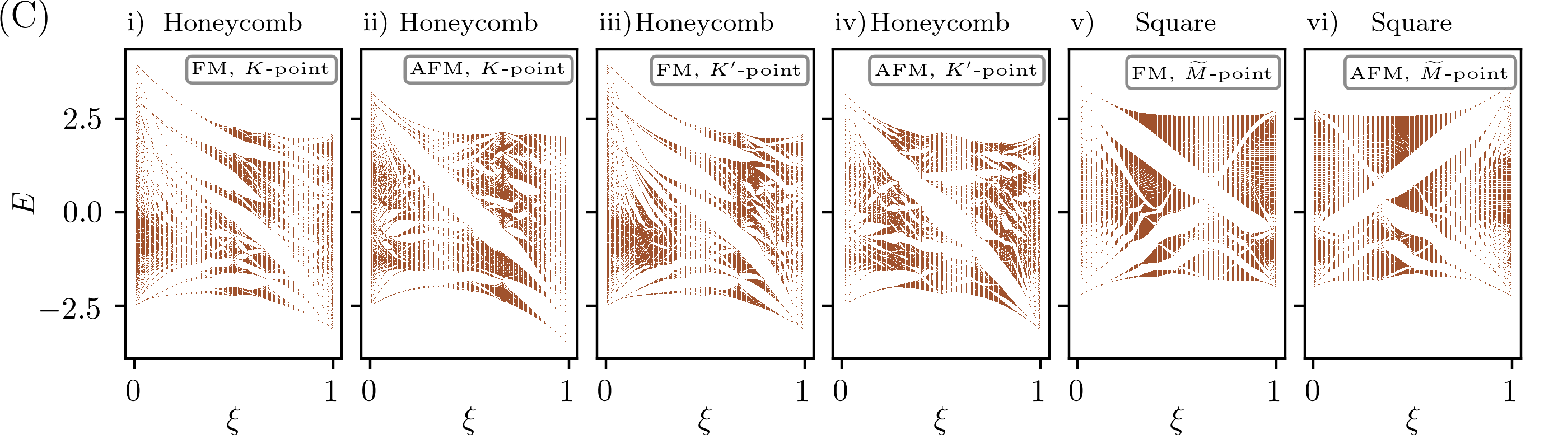}
    \caption{Hofstadter-type spectra for the (A)~honeycomb and (B)~square lattices, at the FMBZ $\Gamma$-points. The values of amplitudes $\tnn$ and $\tnnn$ are indicated in the insets. The green solid lines demonstrate the eigenenergies of the corresponding $\ham_{\text{nnn}}$. The blue solid lines represent the eigenenergies calculated within the stationary perturbation theory (see the text for details).
    (C) The butterflies calculated for (i--iv) honeycomb and (v, vi) square lattices for different magnetic patterns. The tunneling amplitudes are $\tnn = \tnnn = 0.5$, the magnetic phases and points of the FMBZ are specified in the insets.
    (A--C) Everywhere, the horizontal axis is responsible for the magnetic flux, $\xi = p/q$ with $q=197$ and $q = 1, 2, \ldots q-1$, the vertical axis is the eigenenergy.}
    \label{fig:butterflies}
  \end{figure*}

  Figures~\ref{fig:butterflies}A and \ref{fig:butterflies}B demonstrate the eigenenergies of the bulk Hamiltonians, \eqref{eq:Hhc_bulk} and \eqref{eq:Hsq_bulk}, respectively, computed at the $\Gamma$-points, $k'_1 = k_2 = 0$. The spectra are evaluated for the conventional values of the magnetic flux, $\xi \in (0, 1)$, namely $\xi = p/q$ with $q = 197$ and $p = 1, 2, \ldots (q{-}1)$.
  The pair of NN and NNN transitions amplitudes, $(\tnn,\tnnn)$, gradually changes from $(0, 1)$ to $(1, 0)$.

  Consider the extreme situations when one of the amplitudes equals zero. The honeycomb lattice with $\tnn {=} 0$ and $\tnnn {=} 1$ is equivalent to the pair of independent ferromagnetic triangular lattices (\textsl{cf.} \cite{Oh2001}). Similarly, the square lattice can be treated as a pair of independent ferromagnetic square ones (\textsl{cf.} \cite{Hofstadter1976}). In these cases, the eigenstates are at least doubly degenerate.
  On the other hand, the case when $\tnn {=} 1$ and $\tnnn {=} 0$ corresponds to the absence of magnetic field effect and no dependence of the eigenenergies on $\xi$, as it can be clearly seen from the right panels of Figures \ref{fig:butterflies}A and B. The lattices are then equivalent to honeycomb and square ones with the nearest-neighbour hopping only.

  In addition, second from the left panels in Fig.~\ref{fig:butterflies}A and B contain green and blue solid lines representing the highest energetic states of NNN-hopping parts of the bulk Hamiltonians \eqref{eq:bulk_hams} with $\tnnn = \sfrac{3}{4}$ (green lines) and their splitting by the influence of correspoding NN-hopping parts (with $\tnn = \sfrac{1}{4}$)
  treated within the first order of stationary perturbation theory (the sets of blue lines). As can be seen, such approach descibes the effect for small $\tnn$ and weak gauge fields sufficiently well.

  To illustrate the difference between the AFM and FM phases, the Hofstadter-type spectra were calculated for other points in the FMBZ.
  Figure~\ref{fig:butterflies}C shows the butterflies for the (i--iv) honeycomb and (v--vi) square structures computed for $\tnn = \tnnn = \sfrac12$. $K$ and $K'$ points of the hexagonal FMBZ are located at $(2\pi/3, -2\pi/3)$ and vice versa, $\mpoint$ point of the square FMBZ is positioned at $(\pi/2, \pi/2)$. One can easily see that AFM and FM structures generally result in the different butterfly spectra. The interesting aspect of Figures~\ref{fig:butterflies}Ci) and Ciii) is that the spectra are very similar but still are not completely equal (the absence of the exact match can be detected by the direct comparison of the raw output data of the calculations and is poorly visible in the butterfly charts).

  Finally, Figures~\ref{fig:butterflies}Cv) and \ref{fig:butterflies}Cvi) exhibit the HBs for FM and AFM square lattices at $\mpoint$-point, which was chosen instead of the conventional $M$-point ($k'_1 = k_2 = \pi$) due to the following peculiarity of the square lattice bulk Hamiltonian \eqref{eq:Hsq_bulk}: $k_2$ equal to $\pm\pi$ eliminates the nearest-neighbour hopping part of the Hamiltonian \eqref{eq:Hsq_bulk} as the matrix
  $\bigl( (0, 1+\e^{-i k_2}), (0, 0)\bigr)$
  can be, after a certain algebra, factorized out in the tensor product proportional to $\tnn$.

  Our results can be compared to those of Otaki and Fukui \cite{Otaki2019}, where the authors consider  gradual variation of Hofstadter butterflies arising in the 2D generalization of Su-Schrieffer-Heeger model. In constrast with Fig.~\ref{fig:butterflies}, their HB-spectra retain reflection symmetry with respect to $E = 0$. Another point of connection is the symmetry enclosed in equations~\eqref{eq:Hams_k_minusk}, the analogy of which is reported in the work \cite{Otaki2019}. As well, the results resembling those presented in our paper can be found in the work by Hasegawa and Kohmoto \cite{Hasegawa2013}, where Hofstadter butterflies distortions are examined in twisted bilayer graphene, where such features as splitting of the highest energetic states highlighted by the blue and green lines in Fig.~\ref{fig:butterflies} were reported.
\end{section}

  \begin{section}{Conclusion \label{sec:conclusion}}

  We have developed the theory describing both the ferromagnetic and antiferromagnetic patterns with nearest and next-to-nearest neighbours hoppings in the honeycomb and square lattices. The gauge field has been assumed to alter only the next-to-nearest neighbour tunnelings, the additional phase has not been acquired during transitions to nearest neighbours.
  The major finding of the study was that a Hofstadter butterfly can arise in the AFM structures with zero total gauge field flux through the lattice.

  We have shown that the AFM and FM Hamiltonians and butterflies coincide in the $\Gamma$-points and confirmed the differences for other points in the first magnetic Brillouin zone. Accidental similarities between the butterflies have been also disclosed. For several cases, first order perturbation theory has been applied and demonstrated good agreement with the exact calculations for small magnetic fields.

  This study substantially enriches  the class of the systems where the fractal spectrum may be realized, and paves the way to the search of materials with AFM order, where the Hofstadter Butterflies may be observed.

\end{section}

  \ack
The authors acknowledge support from Russian Science Foundation (project No. 18-72-10110). The work of IAS was also supported by Icelandic Science Foundation (Project "Hybrid Polaritonics"). YSK is grateful to Dr. A. Nalitov for the useful discussion.

  \begin{appendix}
  \begin{section}{Intermediate and final Hamiltonians}{\label{sec:App_expilicHams}}
    In the real space along $\vec{a}_1$, the Hamiltonians expressed in terms of $k_2$ and $\nnew$ are
      \begin{subequations}
        \begin{align}
          &\Hhc[\text{k}1]' =
          \nonumber \\
          &= \sum\limits_{\nnew} \left\{
          \tnn \left[
          |\nnew\rangle \langle\nnew| \otimes
          \left(
          \iq \otimes
          \begin{pmatrix}
            0 & \e^{-i k_2} \\
            1 & 0
          \end{pmatrix}
          +
          \ip{Q}
          \otimes
          \begin{pmatrix}
            0 & 1 \\
            0 & 0
          \end{pmatrix}
          \right) \right. \right. 
          + \left. 
          |\nnew+1\rangle \langle\nnew| \otimes
          \ipp{2Q}
          \right]
          \nonumber \\
          &+
          \tnnn \left[
          |\nnew+1\rangle \langle\nnew| \otimes
          \ipp{Q} \otimes \cq_{Q-1}
          + \e^{-i k_2}
          |\nnew\rangle \langle \nnew+1 | \otimes
          \ippt{Q} \otimes \cq_{Q-1}
          \right. 
          \nonumber \\
          &+ |\nnew\rangle \langle\nnew| \otimes
          \begin{pmatrix}
            \OO & & & & \\
            \cq_0 & \ddots & & & \\
            & \cq_1 & \ddots & & \\
            & & \ddots & \ddots & \\
            & & & \cq_{Q-2} & \OO
          \end{pmatrix}_{2Q}
          \nonumber \\
          &+
          \left. \left. 
          \e^{-i k_2} |\nnew\rangle \langle\nnew|
          \begin{pmatrix}
            \dq_0 & \cq_0 & & & \\
            & \dq_1 & \cq_1 & & \\
            & & \ddots & \ddots & \\
            & & & \ddots & \cq_{Q-2} \\
            & & & & \dq_{Q-1}
          \end{pmatrix}_{2Q}
          \right] \right\} + \hc \label{eq:Hhc_k2(2)}
        \end{align}
        and
        \begin{align}
          \Hsq[\text{k}1]' &= \sum\limits_{\nnew} \left\{
          \tnn \left[
          |\nnew\rangle \langle\nnew| \otimes
          \left(
          \iq \otimes
          \begin{pmatrix}
            0 & \e^{-i k_2} \\
            1 & 0
          \end{pmatrix}
          +
          \ip{Q} \otimes
          \begin{pmatrix}
            0 & 1 + \e^{-i k_2} \\
            0 & 0
          \end{pmatrix}
          \right) \right. \right. \nonumber \\
          &+ \left.
          |\nnew+1\rangle \langle\nnew| \otimes
          \ipp{Q} \otimes
          \begin{pmatrix}
            0 & 1 + \e^{-i k_2} \\
            0 & 0
          \end{pmatrix}
          \right] \nonumber \\
          &+ \tnnn \left[
          |\nnew\rangle \langle\nnew| \otimes
          \left(
          \ip{Q} + \e^{-i k_2}
          \begin{pmatrix}
            \fq_{0} & & & \\
            & \fq_{1} & & \\
            & & \ddots & \\
            & & & \fq_{Q-1}
          \end{pmatrix}_{2Q}
          \right)
          \right. \nonumber \\
          &+ |\nnew+1\rangle \langle\nnew| \otimes
          \ipp{Q}
          \otimes \sigmao
          \biggr] \biggr\} + \hc
          \label{eq:Hsq_k2(2)}
        \end{align}
      \end{subequations}
      Here, all the empty positions in the matrices indicate zeros, $\OO$ is the $2{\times}2$ zero matrix, the indices $Q$ and $2Q$ explicitly show the dimesions of the corresponding matrix. $\iq$ is the $Q$-dimensional identity matrix, $\ip{S}$ and $\ipp{S}$ are auxiliary $S$-dimensional matrices:
      \begin{equation}
        \ip{S} =
        \begin{pmatrix}
          0 & & & \\
          1 & \ddots & & \\
          & \ddots & \ddots & \\
          & & 1 & 0
        \end{pmatrix}_{S},
        \qquad
        \ipp{S} =
        \begin{pmatrix}
          0 & & 1 \\
          & \ddots & \\
          & & 0
        \end{pmatrix}_{S},
      \end{equation}
      with $S$ being equal to either $Q$ or $2Q$. The superscript $\mathsf{T}$ implies matrix transposition.
      Performing the Fourier transform over $\nnew$ coordinate responsible for the supercell position, we finally get the explicit expressions for the Hamiltonians in the reciprocal space:
      \begin{subequations}{\label{eq:bulk_hams}}
        \begin{align}
          \Hhc[\text{bulk}] &=
           \, \tnn \left[
          \iq \otimes
          \begin{pmatrix}
            0 & \e^{-i k_2} \\
            1 & 0
          \end{pmatrix} +
          \ip{Q} \otimes
          \begin{pmatrix}
            0 & 1 \\
            0 & 0
          \end{pmatrix} +
          \e^{-i k'_1}
          \ipp{2Q}
          \right] \nonumber \\
          &+ \tnnn
          \left[
          \begin{pmatrix}
            \OO & & & & \\
            \cq_0 & \ddots & & & \\
            & \cq_1 & \ddots & & \\
            & & \ddots & \ddots & \\
            & & & \cq_{Q-2} & \OO
          \end{pmatrix}_{2Q} +
          \e^{-i k_2}
          \begin{pmatrix}
            \dq_0 & \cq_0 & & & \\
            & \dq_1 & \cq_1 & & \\
            & & \ddots & \ddots & \\
            & & & \ddots & \cq_{Q-2} \\
            & & & & \dq_{Q-1}
          \end{pmatrix}_{2Q}
          \right. \nonumber \\
          &+ \e^{-i k'_1}
          \ipp{Q} \otimes \cq_{Q-1}
          + \e^{i (k'_1 - k_2)}
          \ippt{Q} \otimes \cq_{Q-1}
          \biggr] + \hc
            \label{eq:Hhc_bulk}
        \end{align}
        and 
        \begin{align}
          \Hsq[\text{bulk}] &=  \, \tnn \left[ \iq \otimes
          \begin{pmatrix}
            0 & \e^{-i k_2} \\
            1 & 0
          \end{pmatrix} +
          \ip{Q} \otimes
          \begin{pmatrix}
            0 & 1 + \e^{-i k_2} \\
            0 & 0
          \end{pmatrix}
          + \e^{-i k'_1}
          \ipp{Q} \otimes
          \begin{pmatrix}
            0 & 1 + \e^{-i k_2} \\
            0 & 0
          \end{pmatrix}
          \right] \nonumber \\
          &+ \tnnn \left[
          \ip{Q} \otimes \sigmao
          + \e^{-i k_2}
          \begin{pmatrix}
            \fq_{0} & & & \\
            & \fq_{1} & & \\
            & & \ddots & \\
            & & & \fq_{Q-1}
          \end{pmatrix}_{2Q} +
          \e^{-i k'_1}
          \ipp{Q}
          \otimes \sigmao
          \right] + \hc
          \label{eq:Hsq_bulk}
        \end{align}
      \end{subequations}
  \end{section}
\end{appendix}

  \bibliographystyle{unsrt}
  \bibliography{bibitems}

\end{document}